\documentstyle[aps,prl,epsf]{revtex}  
\begin{document}
\draft


\twocolumn[\hsize\textwidth\columnwidth\hsize  
\csname @twocolumnfalse\endcsname              

\title{ Color-Octet Fraction in $J/\psi$ Production and Absorption}

\author{ Cheuk-Yin Wong}

\address{ Oak Ridge National Laboratory, Oak Ridge, TN 37831}

\author{ Chun Wa Wong}

\address{University of California, Los
Angeles, CA 90095-1547 }


\maketitle
 
\begin{abstract}
The cross section between a $c\bar c$ pair and a nucleon is small and
sensitive to the $c$-$\bar c$ separation if the pair is in a
color-singlet state, but very large and insensitive to the separation
if it is in a color-octet state. We use this property in an absorption
model involving both color components to deduce the color structure of
$c\bar c$ pairs produced in $p(B)A \rightarrow \psi X$ reactions. Our
analysis shows that the NA3, NA38 and E772 data are not inconsistent
with the theoretical picture that color-octet and color-singlet
precursors are produced in roughly equal proportions if the produced
color-singlet precursors are pointlike and transparent. However, if
the color-singlet precursors are not transparent but have a cross
section of a few mb, these data do show a definite preference for a
larger fraction of color-singlet precursors. In either case, the
color-octet fraction increases with $x_{{}_F}$, approaching unity as
$x_{{}_F}$ becomes large.

\end{abstract}


   ]  

\pagebreak

\narrowtext

\section{ Introduction } 

There has been much recent interest in the mechanisms of heavy
quarkonium production.  Bodwin, Braaten, and Lepage \cite{Bod95} have
developed a factorization formalism based on nonrelativistic quantum
chromodynamics (NRQCD) for very massive quarks, a formalism that
allows a systematic calculation of inclusive $J/\psi$ production cross
sections.  The formalism accounts for the production of both
color-singlet (C1) and color-octet (C8) $c\bar c$ precursor states
that will evolve into C1 quarkonium states.  It has been used to study
many heavy quarkonium production processes
\cite{Bra95,Tan96,Tan96a,Ben96,Bra96,Cho96,Fle95,Cac96,Amu96,Gup96,Gup96a,Sle96}.

In NRQCD, production amplitudes are expanded in powers of both the
strong coupling constant $\alpha_s$ and the velocity $v$ of the heavy
quark. For hadroproduction of quarkonia at fixed-target energies of
several hundred GeV, the lowest (called hereafter the ``leading'')
order in $J/\psi$ production turns out to be $\alpha_s^3 v^3$ for C1
precursors, and $\alpha_s^2 v^7$ for C8 precursors. Theoretical
analyses have shown that in these leading orders, the total $J/\psi$
production comes from C8 and C1 precursor states in roughly equal
proportions \cite{Tan96,Tan96a,Ben96,Foot1}.  

However, for hadroproduction of $J/\psi$, $\psi'$ and $\chi$ with low
$p_t$ at fixed-target energies, the calculated lowest-order results of
this double expansion \cite{Tan96,Tan96a} seem to disagree with the
observed polarization and production rates of $J/\psi$ and
$\chi_{1,2}$. Although one can adjust input parameters to fit the
observed production rates, the discrepancy with the polarization data
remains \cite{Ben96}.  This seems to indicate a need for higher-order
quarkonium production mechanisms at these energies
\cite{Tan96,Tan96a,Ben96}.

One of the important parameters that characterize the nature of these
quarkonium production processes is the color-octet fraction at
production. We would like to point out in this paper that this
information can be extracted from the observed nuclear suppression of
$pA$ or $BA \rightarrow \psi X$ cross sections. The possibility arises
because the produced C8 precursors are expected to be absorbed much
more strongly than C1 precursors
\cite{Lip75,Don92,Low75,Nus75,Gun77,Lev81,Lan87,Dol92,Duc93,Won96b,Fra94}.
This possibility is realized by generalizing the absorption model
\cite{And77,Ger88,Won94,Won96,Won97}, in Sec. II, to handle these two
color components. The color dependence of $c\bar c$-$N$ cross sections
are then reviewed in Sec. III to provide a theoretical background
against which the analysis of the available experimental data for low
$p_t$ $J/\psi$ production at fixed-target energies
\cite{Bad83,Ald91,Lou95} will be made, in Sec. IV, using our
two-component absorption model.

Our analysis shows that the data are not inconsistent with the
theoretical picture that C8 and C1 precursors are produced in roughly
equal proportions if the C1 precursors are produced in pointlike
transparent or noninteractive states. However, when freed from these
prevalent theoretical prejudices, the available data do show a
definite preference for a larger fraction of C1 precursors if they are
produced, and are propagating, in states that are significantly
absorbed by the nuclear medium. In either case, the C8 fraction
increases with the Feynman $x_{{}_F}$, approaching unity as $x_{{}_F}$
becomes large.

Additional implications of our models are briefly discussed, and the
need for more experimental absorption data is noted, in the concluding
Sec. V.

\section{ A generalized absorption model with two color components} 

In NRQCD \cite{Bod95}, dynamical processes in NRQCD are controlled by
various time scales: (1) the quark-antiquark production time $1/M$
where $M$ is the $c$ quark mass, (2) the time for orbital motion in
quarkonium $1/Mv \approx r \approx \sqrt{1/M\Lambda_{QCD}}$, where $r$
are the characteristic spatial extension of the quarks in the $c\bar
c$ pair and $\Lambda_{QCD}$ is the QCD confinement scale, and (3)
$1/Mv^2 = 1/\Lambda_{QCD}$ \cite{Bod95} for either the characteristic
time for the $c\bar c$ pair to be blown up from a point to quarkonium
size, or equivalently the QCD confinement time.  The C8 precursor will
eventually hadronize into C1 $J/\psi$ mesons by color neutralization
through the absorption or emission of soft gluons by the end of the
strong-interaction time.

The traditional understanding is that this color-neutralization
process takes place over a much longer nonperturbative QCD time scale
of about $1/\Lambda_{QCD} \approx 0.5$ fm/c in the $c\bar c$ rest
frame. In this frame, the longitudinal spacing between target nucleons
in a $pA$ reaction is $d/\gamma(x_{{}_F})$, where $d=2$ fm is the
internucleon spacing in a nucleus at rest and $\gamma (x_{{}_F})$ is
the relativistic energy/mass ratio of the moving target nucleons,
\begin{eqnarray}
\gamma(x_F)=&& \sqrt{s_{NN}}\biggl \{
\sqrt{m_{J/\psi}^2+x_F^2s_{NN}/4+p_{t,J/\psi}^2}\nonumber\\
&&+x_F\sqrt{(s_{NN}-4m_N^2)/4} \biggr \}/(2m_{J/\psi}m_N) \, ,
\end{eqnarray}
where $p_{t,J/\psi}$ is the transverse momentum of the produced
$J/\psi$, and $<p_{t,J/\psi}^2>=1.26$ GeV$^2$\cite{Ant92}.  Thus, the
dynamics of $J/\psi$ propagation after production in nuclei is further
controlled by the passage time $d/\gamma\beta=d/(\gamma^2 -1)^{1/2}$
the next target nucleon takes to meet the produced $c\bar c$ pair.
Since the value of $\gamma(x_{{}_F})$ can be large in high-energy $pA$
collisions (about 15 at $x_F=0$ when the $NN$ c.m. energy is
$\sqrt{s_{NN}} = 30$ GeV), one finds
$d/\gamma(x_{{}_F})\beta(x_{{}_F}) << 0.5$ fm/c for $x_{{}_F}>0$ at
fixed-target energies of several hundred GeV. Therefore, for $pA$
collisions in fixed-target experiments, many of the collisions between
target nucleons and the produced $(c \bar c)_8$ pair with $x_{{}_F}>0$
are expected to take place before its color is neutralized. This is
particularly true at higher energies where the Lorentz contraction is
stronger.

The collisions of this C8 $c\bar c$ pair with target nucleons at high 
energies have been studied earlier by Kharzeev and Satz \cite{Kha93}. 
They have argued that these collisions do not lead to absorption 
(the eventual breakup of the $c\bar c$ system).  They assume instead 
that the pair will stay together as it traverses the medium,
suffering only quasi-elastic scatterings caused by stretchings of the
$(c\bar c)_8$ string that shift the same integrated production cross
section to lower $x_{{}_F}$. To account for the nuclear suppression
shown in the data, they appeal to the idea of gluon shadowing,
i.e. the assumption of a nuclear modification of the gluon density of
target nucleons that depends only on the fractional momentum $x_2$
carried by the target partons \cite{Kha93,Gup92}.

We would like to describe here a very different picture of $J/\psi$
suppression in nuclei based on a generalization of the standard
absorption picture of \cite{And77,Ger88,Won94,Won96,Won97}. A
precursor can remain in the same precursor state after colliding with
a target nucleon, but its transformation into other precursors through
the exchange of a Pomeron or a hard gluon cannot in general be
avoided. The only exception is for C1 precursors in the pointlike, or
color transparency, limit, a situation we shall discuss further below.
The $c\bar c$ precursor could still stay close together, but its
future fate in the absence of further collisions is already determined
in this precursor representation of states. When the precursor remains
in its original precursor state after scattering, we have elastic
scattering. All other scattering processes contribute to the reaction
cross section $\sigma_r$.

We begin by considering the hard scattering between a parton of the 
projectile nucleon and a parton of a target nucleon inside a nucleus 
with $A$ nucleons, a hard scattering that produces both C1 and C8 
precursor $(c\bar c)$ pairs which will evolve into various quarkonium 
and open-charm meson states.  The probability element for 
precursor production by the collision at a
target nucleon at ${\bf r}_{{}_A}=({\bf b}_{{}_A},z_{{}_A})$ is
$$\rho({\bf b}_{{}_A},z_{{}_A}) d{\bf b}_{{}_A} ~ dz_{{}_A}\,,$$ 
where the density distribution is normalized by 
$$ \int \rho({\bf r}_{{}_A}) d{\bf r}_{{}_A}=1.$$ A produced precursor
will collide with target nucleons along its path with a $(c\bar
c)$-$N$ reaction cross section of $\sigma_{ r}$.  The probability of
the precursor colliding with a target nucleon is therefore
$$T_{A>}({\bf b}_{{}_A},z_{{}_A})\sigma_{r}\,,$$ where
\begin{eqnarray}
T_{A>}({\bf b}_{{}_A},z_{{}_A})=\int_{z_{{}_A}}^\infty
\rho({\bf
b}_{{}_A},z_{{}_A}') dz_{{}_A}',
\end{eqnarray}
and $T_{A>}({\bf b}_{{}_A},-\infty)= T_{{}_A}({\bf b}_{{}_A})$, the
usual thickness function.  Thus, the probability for the precursor to
collide with $n$ target nucleons and miss the other $(A-1)-n$ target
nucleon is
\begin{eqnarray}
{A-1 \choose n}
[T_{A>}({\bf b}_{{}_A},z_{{}_A}) \sigma_r ]^n [1-T_{A>}({\bf
b}_{{}_A},z_{{}_A})\sigma_r]^{(A-1)-n}\,.
\end{eqnarray}

  After the precursor has collided with the target nucleons, the
precursor will be in different degrees of woundedness. We denote by
$S_n$ the probability of finding the $J/\psi$ precursor (which will
eventually evolve into a $J/\psi$ at the end of the strong-interaction
time) after colliding with $n$ target nucleons.  The meson production
cross section in a $pA$ collision with a nuclear target of mass number
$A$ can then be related to the production cross section in
nucleon-nucleon collision by 
\begin{eqnarray}
&&{d\sigma_{J/\psi}^{pA}/dx_F \over A d\sigma_{J/\psi}^{NN}/dx_F} =
\int \rho({\bf b}_{{}_A},z_{{}_A}) d{\bf
b}_{{}_A} ~ dz_{{}_A} \sum_{n=0}^{A-1} S_n 
{ A-1 \choose n}
\nonumber\\
&& \times 
[T_{A>}({\bf b}_{{}_A},z_{{}_A}) \sigma_r ]^n [1-T_{A>}({\bf
b}_{{}_A},z_{{}_A})\sigma_r]^{(A-1)-n}\,.
\end{eqnarray}
By integrating over $z_{{}_A}$ and extending the above considerations 
to include both C1 ($i=1$) and C8 ($i=8$) components, we obtain
the $x_{{}_F}$-dependent nucleus to nucleon yield ratio per nucleon for 
the quarkonium under consideration:

\begin{eqnarray}
\label{eq:sum}
R (pA/NN, x_{{}_F}) && \equiv 
{d\sigma_{J/\psi}^{pA}/dx_F \over A d\sigma_{J/\psi}^{NN}/dx_F}
\nonumber \\ && =
\sum_{i=1,8} f_i (x_{{}_F}) 
\sum_{n=0}^{A-1} S_{i \, n} R_{i \, n}(A)\,,  
\end{eqnarray}
where $f_i(x_{{}_F})$ is the $(c\bar c)_i$ fraction normalized to 
$f_1 + f_8 = 1$, and 

\begin{eqnarray}
&& R_{i \, n}(pA) = 
\int {d{\bf b}_{{}_A} \over \sigma_{i \, r}} 
\sum_{m=0}^n  
{ A-1 \choose n} \nonumber\\ 
&&
\times 
{n \choose m}
{(-1)^m \over A-n+m}
\biggl [ 1 - \biggl ( 1 - T_{{}_A}({\bf b}_{{}_A}) \sigma_{i \, r} \biggr )
^{A-n+m} \biggr ] \,.
\end{eqnarray}
This is our generalized absorption model.

It is clear that when a $J/\psi$ precursor produced at a nucleon site
passes through the rest of the target nucleus without further
collision, it will evolve into a final-state $J/\psi$ as if it had
been produced in a $pN$ collision in free space: a C1 precursor will
evolve into a $J/\psi$, while a C8 precursor $(c\bar c)_8$ will evolve
into a $J/\psi$ with the additional absorption or emission of a soft
gluon after a relatively long QCD color neutralization time that is
still short compared to electromagnetic interaction times.  Because of
this, $S_{i \, 0}$ is unity by definition.  Furthermore, $f_i$ are the
actual color fractions right after production at the production site
of a target nucleon.

A minor complication should now be mentioned. In addition to the
direct production considered so far, the experimental detector also
counts $J/\psi$ particles that come indirectly from radiative decays
of excited quarkonium states, particularly the $\chi_{1, 2}$ mesons.
These indirect contributions can simply be added to the direct
contribution in both the numerator and the denominator differential
cross sections that make up the yield ratio $R$ in Eq. (\ref
{eq:sum}). Equivalently, as we choose to do from now on, we can
re-define our precursor states so that they include both direct and
indirect $J/\psi$ mesons that will enter the experimental detector.

The original collision at a production site produces precursors not
only for the final $J/\psi$, but also for all other permissible
hadronic final states not included in the experimental yield for
$J/\psi$ production.  Hence these other precursors do not contribute
to our model formula when there is no further collision at the target
nucleus.

Let us consider next a precursor that suffers one or more collisions
in the target nucleus after production. At the end of all these
collisions, the original $J/\psi$ precursor will be transformed into
precursors for all possible final hadronic states including $J/\psi$,
with a total probability of 1. At the same time, other precursors
different from the $J/\psi$ precursor, all produced at the original
target nucleon site, will be changed into $J/\psi$ precursors with
some finite probabilities. The normalized probability $S_{i \, n}$ for
$n \ge 1$ is just the population of $J/\psi$ precursors present after
$n$ collisions with target nucleons, normalized to a $J/\psi$
precursor population of $S_{i \, 0} = 1$ for precursors that escape
any hit. Containing contributions from all precursors produced at the
production site, it describes the probability of {\it recovering} a
$J/\psi$ precursor after $n$ precursor-nucleon collisions.

These recovery probabilities are relatively complicated quantities
that contain the effects of available phase space and of coherent
coupled-channel dynamics \cite{Won96b}. A simple assumption one can
make is that on the average a certain fraction of $\sigma_{i\,r}$ is
recoverable, while the remainder, denoted the effective absorption
cross section $\sigma_{i\,{\rm abs }}$ in nuclei, is irretrievably
lost.  Using this fractional $\sigma_{i\, {\rm abs }}$ in our
formulas, we should now set all recovery probabilities $S_{i \, n}$
for $n \ge 1$ to zero, because precursors are now, by definition,
irretrievably lost after each hit by the effective $\sigma_{i\, {\rm
abs }}$. Generalizing to nucleus-nucleus ($BA$) collisions, we obtain
the following equation for the $x_{{}_F}$-dependent nucleus-nucleus to
nucleon-nucleon yield ratio per target nucleon per projectile nucleon
for $J/\psi$ production:

\begin{eqnarray}
\label{eq:yratio}
R (BA/NN, x_{{}_F})
 = \sum_{i=1,8} f_i (x_{{}_F}) R_i(BA)\,,  
\end{eqnarray}
where $f_i(x_{{}_F})$ is the $(c\bar c)_i$ fraction normalized to 
$f_1 + f_8 = 1$, and 

\begin{eqnarray}
\label{eq:xsec}
R_i(BA) = && \int {d{\bf b}_A \over A\sigma_{i \,{\rm abs}}} 
{d{\bf b}_B \over B\sigma_{i \,{\rm abs}}}  \nonumber \\
&& \times \biggl \{ 1 - \biggl ( 1- T_B ({\bf b}_B) 
\sigma_{i \,{\rm abs}} \biggr )^B \biggr \} \nonumber \\
&& \times \biggl \{ 1 - \biggl ( 1- T_A ({\bf b}_A) 
\sigma_{i \,{\rm abs}} \biggr )^A \biggr \} \, .
\end{eqnarray}
This is just the familiar ``simple'' absorption model, now generalized 
to handle two color components. The absorption cross sections that appear 
are effective values in the nuclear medium involving precursors not at 
the moment of their production, but when they hit the next nucleon in 
the colliding nuclei.

In generalizing the $pA$ result of Eq.\ (\ref{eq:sum}) to Eq.\
(\ref{eq:xsec}) for $BA$ collisions, we have made the implicit
assumption that the absorption of the precursor of $J/\psi$ due to its
collision with produced soft particles is not important in $BA$
collisions.  This is because the average relative kinetic energy
between the produced particles and the precursors of $J/\psi$ is
smaller than the threshold energy (about 640 MeV) for the precursor to
breakup.  It is further supported by comparing experimental $pA$ and
$AB$ data \cite{Won97}.

From the perspective of our generalized absorption picture, the model
of Ref.\ \cite{Kha93} contains only elastic scattering of precursors
and no absorption at all. With no absorption present, the authors are
forced to introduce another source of absorption based on gluon
shadowing.

Gluon shadowing describes a change in the momentum distribution of a 
parton in a nucleon in the target as compared to that in free space. 
The momentum distribution of a projectile parton is also changed 
because of the loss of initial energy due to collisions before the 
hard scattering at the production site.  These shadowing effects are real, 
and they should be included in a complete theory.  However, they appear
to be small, as evidenced by the weak dependence of the charm yield per
nucleon on the target mass number $A$ in $pA$ collisions given by
$A^{1.00\pm 0.05\pm 0.02}/A$ for $x_{{}_F}$ from 0.05 to 0.4
\cite{Alv93}.  Therefore, we shall not include them in our analysis.

\section{ Color-dependence of 
$({\lowercase{c}}\bar {\lowercase{c}})$-$N$ cross sections}

For the absorption cross sections needed in Eq. (\ref {eq:yratio}), we
rely conceptually on the fact that high-energy hadron-hadron cross
sections are dominated by Pomeron exchange \cite{Lip75,Don92}.  In the
Two-Gluon Model of the Pomeron (TGMP) studied by Low, Nussinov and
others \cite{Low75,Nus75,Gun77,Lev81,Lan87,Dol92,Duc93,Won96b}, the
flavor dependence of the total cross sections is a size-dependent
effect arising from the color separation in colorless hadrons. The
total hadron-nucleon cross section can be expressed as $T_1 - T_2$,
where $T_n$ is the contribution in which the two exchanged gluons
interact with $n$ particles (here quarks) in the projectile. The cross
section vanishes if one of the colliding hadrons shrinks to a point,
because in this limit $T_2 = T_1$. In this point limit, the hadron
cannot even scatter into intermediate C8 states by single gluon
exchange because it is color neutral. Thus pointlike C1 precursors are
transparent in the nuclear medium, with zero total cross section when
the colliding energies are sufficiently high so that meson-exchange
contributions become unimportant. This phenomenon of ``color''
transparency is the transparency of pointlike colorless hadrons in a
nuclear medium of large colorless nucleons. (For a recent review of
color transparency, see \cite{Fra94}.)

If the C1 precursors are produced in pointlike states, and if the
collision energies are so high that the passage time to the next
target nucleon is too short for such pointlike precursors to grow much
in size, these C1 precursors will be quite transparent as they
propagate in the nuclear medium. Under the circumstances, nuclear
absorption in $J/\psi$ production can only come from the absorptive C8
precursors. This gives us a window for watching C8 precursors in
$J/\psi$ production. It will be interesting to find out the extent to
which this theoretical picture is actually supported by the
experimental data on nuclear suppression.

The cross section is very different for $(q\bar q)_8$-$N$ scattering,
however, as pointed out by Dolej\v s\' i and H\"ufner
\cite{Dol92}. This is because the one- and two-quark contributions now
add together in the form of $T_1 + T_2/8$.  The result is then
insensitive to the $q$-$\bar q$ separation in C8 precursors. It is
also very large, typically of the order of 30-60 mb when a
perturbative propagator is used for gluons with a nonzero effective
mass.  The situation is reminiscent of that in electrodynamics where
the cross section for two equal charges of the same sign is much
larger than the cross section for a dipole made up of two equal but
opposite charges \cite{Dol92}.

Recently, this TGMP for both singlet and octet $(q\bar q)$-$N$
scattering has been studied in detail by one of us \cite{Won96b}. The
main motivation is to understand why the experimental cross sections
for radially excited mesons of much larger sizes are actually close to
one another in value. This unexpected feature can be understood in the
TGMP if the mesons are propagating in an eigenmode with a common eigen
cross section because of strong coupling between them.  In addition, a
detailed model has been fitted in \cite{Won96b} that contains a number
of important refinements: (1) A nonperturbative gluon propagator (the
Cornwall propagator) is used \cite{Cor82,Hal93} with the gluon mass
obtained by fitting the $\pi N$ and $KN$ total cross sections. (2)
Complete meson form factors are used without making the small meson
approximation. (3) For (singlet meson)-$N$ scattering, a
coupled-channel problem \cite{Huf96} is solved using many scattering
channels containing radially excited mesons.  By fitting $NN$ cross
sections and the ratios of (meson-$N$)/$NN$ cross sections, the
extrapolated octet $(c\bar c)_8-N$ total cross section, to be denoted
$\sigma_{8}$ below, turns out to be 48 mb, in agreement with the range
of $30-60$ mb found by Dolej\v s\' i and H\"ufner\cite{Dol92}.  These
color-octet cross sections are quite insensitive to meson size and
flavor contents.

We shall need in our analysis that part of the reaction cross section
denoted in this paper as the effective $\sigma_{8 {\rm abs}}$ in
nuclei.  A $c\bar c$-$N$ collision at high energies can be expected to
cause the $c\bar c$ pair to be broken up, i.e. removed from the
$J/\psi$ channels, with relatively little elastic or quasi-elastic
scatterings. In the nuclear medium, however, this reaction cross
section must be reduced by its fractional recovery in subsequent
collisions. Hence we shall use the estimated theoretical value only
for conceptual guidance, and shall try to find out what the data might
tell us about this cross section.
 
To complete our review of Pomeron exchange cross sections, we should
point out that the color-singlet $(c\bar c)_1$-$N$ total cross section
in free space, to be denoted $\sigma_{1}$, can be estimated in a
number of ways. The model fitted in \cite{Won96b} gives a result of
5-6 mb at $\sqrt{s} = 20$ GeV, but requires an input of the $J/\psi$
meson rms radius, for which we have only theoretical estimates. A
result of at least 2.5 mb at this energy has been calculated by
Kharzeev and Satz \cite{Kha94} from hadron gluon structure functions
in short-distance QCD. A third estimate can be made by converting the
experimental forward $J/\psi$ photoproduction cross section
\cite{Hol85} to a total $\psi-N$ cross section with the help of
vector-meson dominance (VMD), i.e. the idea that the photon actually
contains a small admixture of vector mesons. This gives $\sigma_{1}
\approx 1.8$ mb at $\sqrt{s}=20$ GeV. However, the VMD model is known
to underestimate the $\rho-N$ cross section by about 15\% and the
$\phi-N$ cross section by about 50\% \cite{Bau78}. This could mean
that $\sigma_{1}$ should be larger, perhaps around 2 to 3.5
mb. Although these three estimates are only in rough agreement with
one another, they are all an order of magnitude smaller than
$\sigma_{8}$.

All these estimates are for the ``asymptotic'' total cross section in
one $c\bar c$-nucleon scattering, and without the additional $s$
dependence appropriate to the Pomeron dominance of the cross sections
at high energies \cite{Don92}.  We are interested only in its
absorptive part in nuclei, after the recovery corrections mentioned
previously. There is, in addition, a threshold effect which reduces
the cross section more and more below its asymptotic value the lower
the collision energy \cite{Kha94}.  Hence we shall adopt a more
opportunistic phenomenological approach in choosing $\sigma_{i\, {\rm
abs}}$ in our model analyses.

\section{ The color-octet fraction} 

We are now in a position to extract the C8 fraction $f_8$ from the
experimental cross section or yield for $J/\psi$ production in nuclei
by using Eq.\ (\ref {eq:yratio}). We first analyze the experimental
$x_{{}_F}$ integrated yields as functions of the target mass number
$A$ at fixed-target energy of 800 GeV of the $pA$ data (E772
Collaboration) \cite{Ald91} and at 200 GeV of the combined $pA$ data
(NA3 Collaboration) \cite{Bad83} and $BA$ (nucleus-nucleus) data (NA38
Collaboration) \cite{Lou95}. (The average values of the kinematical
variables in the experimental data at 800 GeV are $<x_{{}_F}> \approx
0.27$ and $<p_t > \approx 0.7$ GeV \cite{Ald91}.)

The results for the C8 fraction $f_8$ obtained in our model analysis
are shown in Fig. 1a as functions of $\sigma_{\rm 8 \, abs}$ for the
color-transparency choice of $\sigma_{\rm 1 \, abs} = 0$. The
associated $\chi^2$ per degree of freedom of the model fit to data are
given in Fig. 1b. We see that the best fit to the E772 data appears at
$\sigma_{\rm 8 \, abs} = 15$ mb, a value that is considerably smaller
than the best theoretical asymptotic value of 48 mb. However, the data
are consistent with a rather wide range of range $\sigma_{\rm 8 \,
abs}$. The fit is noticeably poorer for the 200 GeV data, which show a
preference for much smaller values of $\sigma_{\rm 8 \, abs}$. This is
probably only partially due to the threshold effect mentioned
previously.

At 800 GeV, the extracted C8 fraction $f_8$ at best fit is about 0.8, 
but the fraction decreases with increasing $\sigma_{\rm 8 \, abs}$, 
being about 0.55 at $\sigma_{\rm 8 \, abs} = 30$ mb. The results at 200 
GeV are noticeably smaller, being usually below 0.5.

It is interesting to compare our results with the information on the
C8 fraction at production deduced from analysis of production data on
nucleon targets. A theoretical analysis of the 300 GeV CDF data on the
$\pi N \rightarrow (J/\psi) X$ by Tang and V\"anttinen \cite{Tan96a}
has yielded a theoretical C8 fraction from both direct production and
indirect production (from their Table 1) of $0.20/(0.20+0.14) \approx
0.59$. However, the total theoretical $J/\psi$ production cross
section is only 0.38 of the observed value. (Indirect production comes
from the radiative decays of excited quarkonium states, primarily
$\chi_{1,2}$.)

\epsfxsize=300pt
\includegraphics{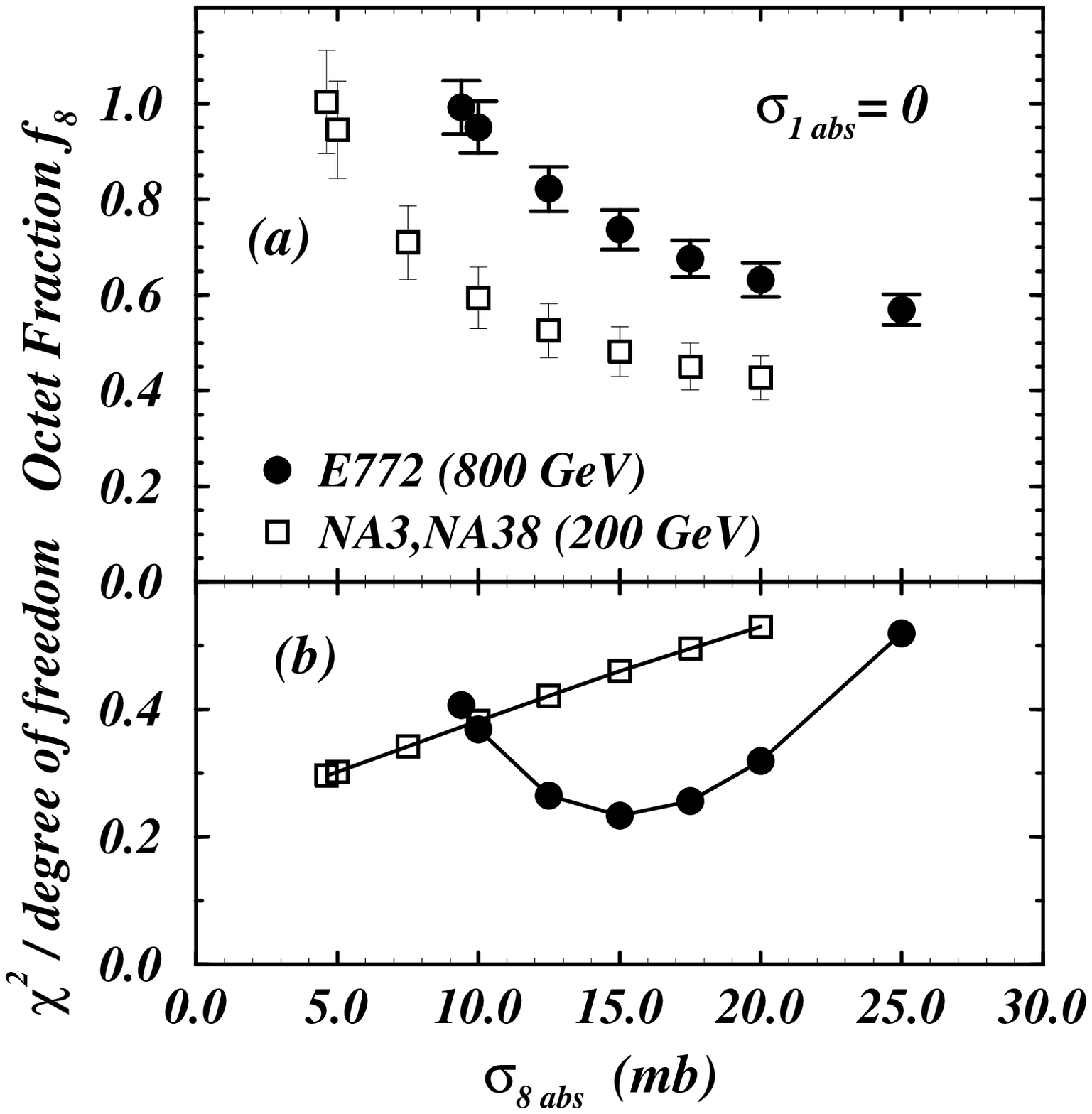}
\vskip 8.5cm
\begin{minipage}[t]{8cm}
\noindent {\bf Fig. 1}.
{(a) The color-octet fraction $f_8$ fitting the experimental
$x_{{}_F}$-integrated yields as a function of the effective
color-octet absorption cross section $\sigma_{\rm 8\, abs}$ when
$\sigma_{\rm 1\, abs}$ is fixed at  zero (color-transparancy limit),
and (b) the corresponding $\chi^2$ per degree of freedom.}
\end{minipage}
\vskip 4truemm
\noindent 

Similar fractions have been obtained by Beneke and Rothstein 
\cite{Ben96} for $pN$ production at 300 GeV. They give a direct C8 
(direct C1) contribution of about 40 (20) \% of the total. This is in 
rough agreement with the direct C8 (direct C1) percentage of 56 (21) \% 
found in \cite{Tan96a}.

Previous analyses of the nuclear suppression data using absorption
models have been based on C1 precursors only, since these analyses
were first performed at a time when production was supposed to be
predominantly C1. With our two-component absorption model, a much
wider range of physical assumptions can be checked against the
experimental data. In particular, we have shown in Fig. 1 that an
absorption model can be constructed that respects the popular
theoretical prejudices that C1 precursors are produced in pointlike
states and tend to be transparent in the colliding nuclear complex,
where C8 precursors are strongly absorbed. The C8 fractions that come
out of this model are quite substantial, in agreement with independent
analyses of hadron production rates in free space.

It is now worth asking if the available nuclear suppression data {\it
require} color transparency. To answer this question, we look for
models with nonzero $\sigma_{\rm 1 \, abs}$. Nonzero absorption for C1
precursors means that they have a substantial size when they hit a
nucleon after production. At fixed target energies, these C1
precursors usually do not have enough time to grow enough in size if
they had been produced pointlike. Thus significant C1 absorption
usually means that these precursors are produced with finite sizes.
\epsfxsize=300pt
\includegraphics{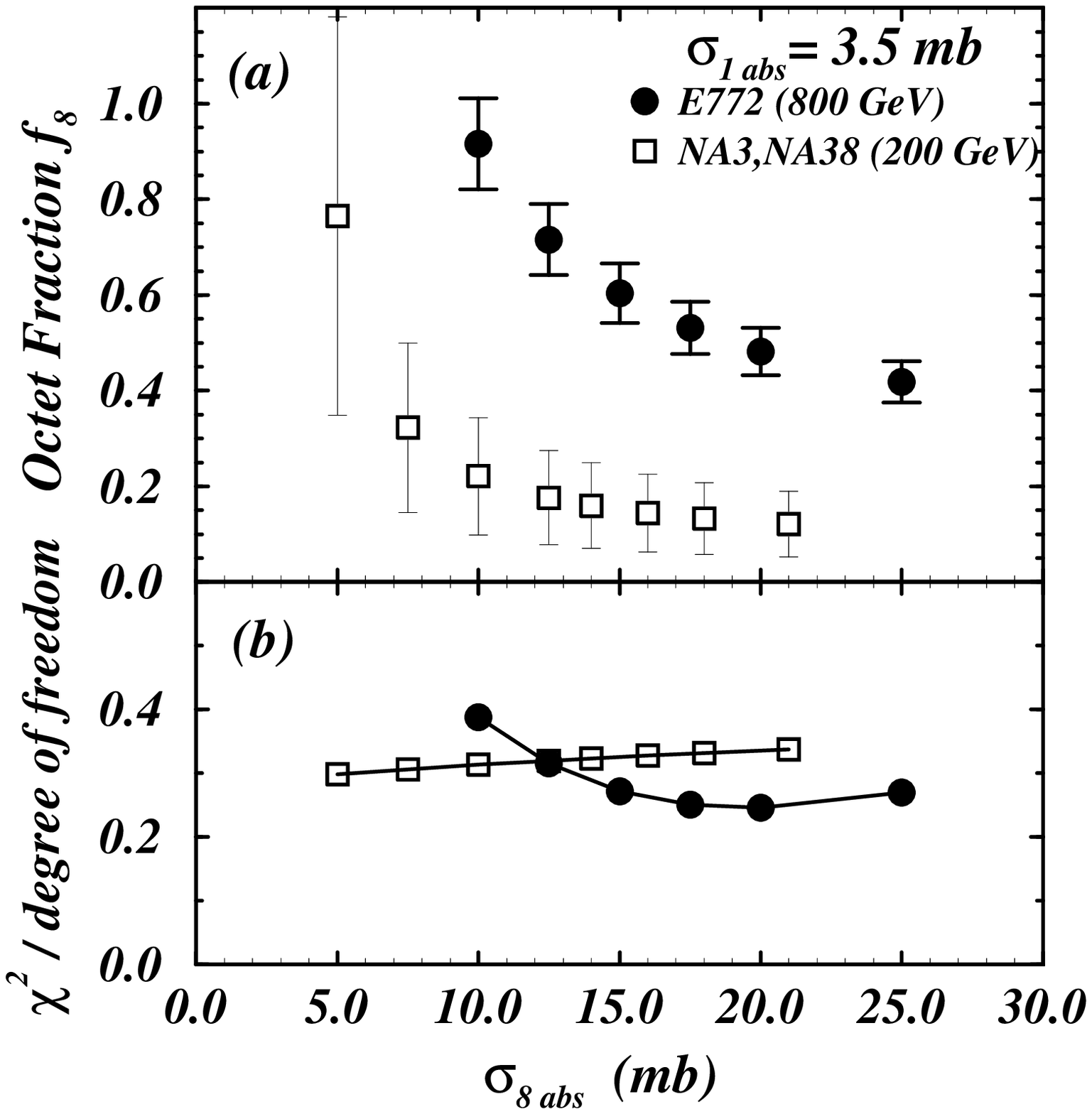}
\vskip 8.5cm
\begin{minipage}[t]{8cm}
\noindent {\bf Fig.2}.  \rm
{Same as Fig. 1 but for $\sigma_{\rm 1\, abs}=3.5$ mb.}
\end{minipage}
\vskip 4truemm
\noindent 

Fig. 2 gives the results for model fitting using $\sigma_{\rm 1 \,
abs} = 3.5$ mb, close to many of the values estimated for the
$J/\psi$-$N$ cross section in free space, as reviewed in Sec.  III. We
see that the fits are comparable to those shown in Fig. 1 for the E772
data, and they are better for the 200 GeV data. Similar fits can be
obtained for the E772 data at $\sigma_{\rm 1 \, abs} = 6.7$ mb, the
best-fit value if $\sigma_{\rm 8 \, abs}$ is fixed at the theoretical
value of 48 mb.  However, the 200 GeV data cannot be fitted well with
this large value of $\sigma_{\rm 1 \, abs}$. One common features of
these models with fairly large $\sigma_{\rm 1 \, abs}$ is that the C1
precursors are now providing a substantial part of the experimental
nuclear suppression. Hence the octet fraction $f_8$ needed is
reduced. Fig. 2 shows that for the 200 GeV data, the extracted $f_8$
is usually less than 0.2. This is much smaller than the octet fraction
found in the theoretical picture of hadronproduction given in
leading-order NRQCD \cite{Tan96,Tan96a,Ben96}.

Our phenomenological analyses seem to show that the available data
alone are not sufficiently discriminating to tell us if the C1
precursors are transparent because they are produced pointlike, or if
they are easily absorbed because they are produced at almost full
size.

\epsfxsize=300pt
\includegraphics{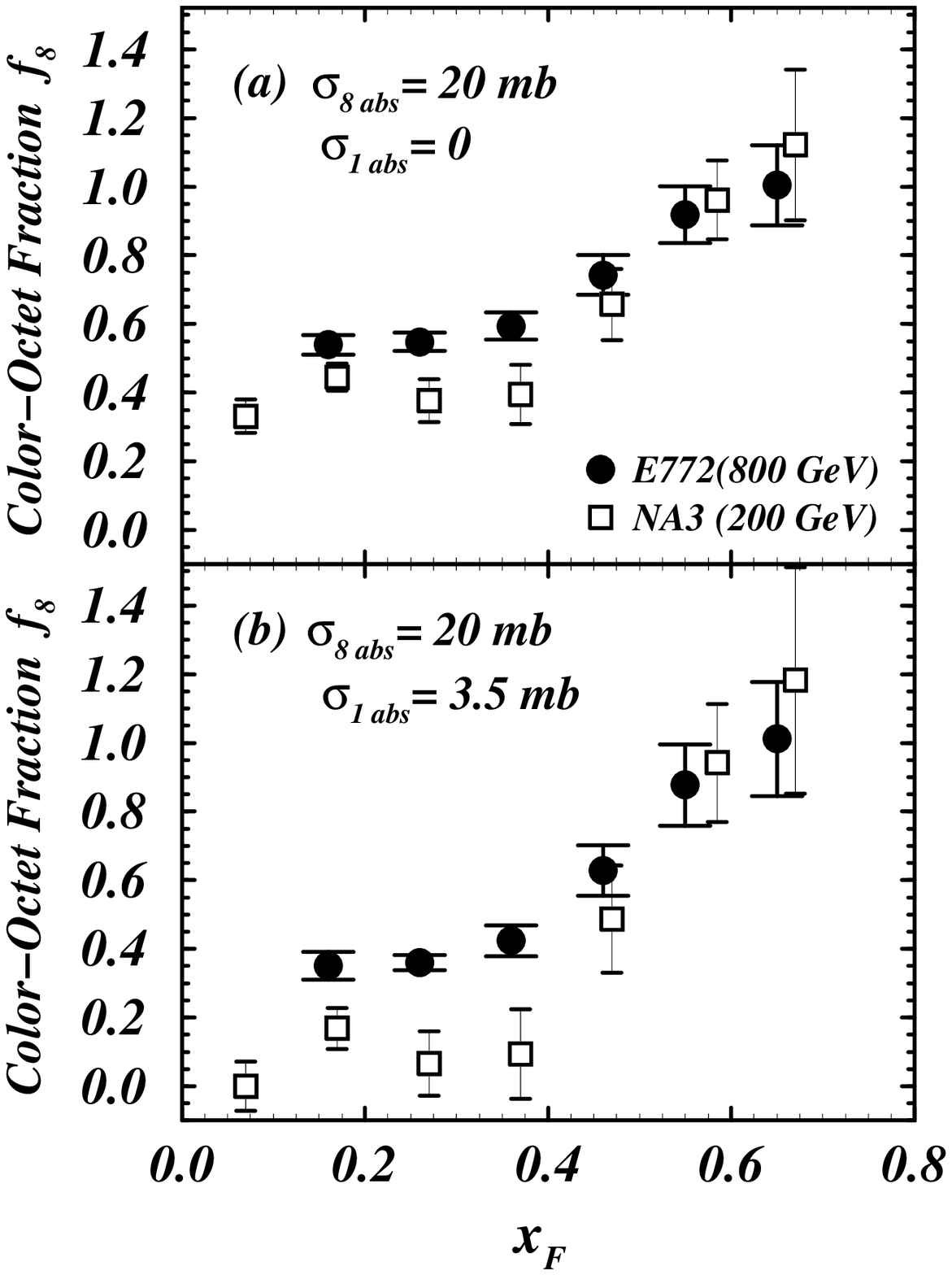}
\vskip 10.9cm
\begin{minipage}[t]{8cm}
\noindent {\bf Fig.3}. 
{The color-octet fraction $f_8$ as a function of
$x_{{}_F}$. (a) is for $\sigma_{\rm 1\, abs}=0$ mb , and 
(b) is for $\sigma_{\rm 1\, abs}=3.5$ mb.}
\end{minipage}
\vskip 4truemm
\noindent 

We next analyze the $x_{{}_F}$-dependent experimental yields for the
$pA$ data using $\sigma_{\rm 8 \, abs} = 20$ mb and $\sigma_{\rm 1 \,
abs} = 0\, (3.5)$ mb. The results for the C8 fraction $f_8$ are given
in Fig. 3a (3b). The error bars shown describe only the uncertainties
from data fitting for the chosen values of $\sigma_{\rm 8 \,abs}$. The
effects coming from the uncertainties of the chosen absorption cross
sections themselves can be seen by comparing the results of Figs. 3a
and 3b, but we should also remember that these figures describe rather
different physical models, one with color transparency and one with
significant C1 absorption.  These figures seem to show that for
$x_{{}_F} > 0.5$, the C8 fraction $f_8$ is rather close to 1, and
seems to scale in $x_{{}_F}$.

To account for the abnormally small yields at large $x_{{}_F}$, Badier
$et~al.$ \cite{Bad83} have to postulate the existence of a new
mechanism of $J/\psi$ production.  We have attributed this phenomenon
instead to the presence of a greater fraction of C8 precursors and to
their strong absorption as they propagate in nuclear matter.  Fig. 3
also shows that below $x_{{}_F} = 0.5$, the extracted $f_8$ fraction
seems to decrease with decreasing collision energy. The decrease is
very dramatic when $\sigma_{\rm 1 \,abs}$ is large.

A tantalizing possibility is that it is not so much the production
mechanisms themselves that are strongly energy dependent, but rather
that the produced precursors at different energies have different
times to evolve before hitting the next nucleon. For example, the
precursors might have been produced predominantly in C8 states, as
suggested by the leading-order NRQCD calculations, but their colors
might have been neutralized in a time-dependent way after production.
It is therefore interesting to plot the deduced C8 fraction against
the passage time $t_p = d/\gamma\beta$ for the two models shown in
Fig. 3.  The results, given in Fig. 4, show that the low-$x_{{}_F}$
points might indeed depend smoothly on $d$, but unfortunately the
points from the two data sets do not overlap so that we cannot
establish a case for this dependence at low-$x_{{}_F}$.
\epsfxsize=300pt
\includegraphics{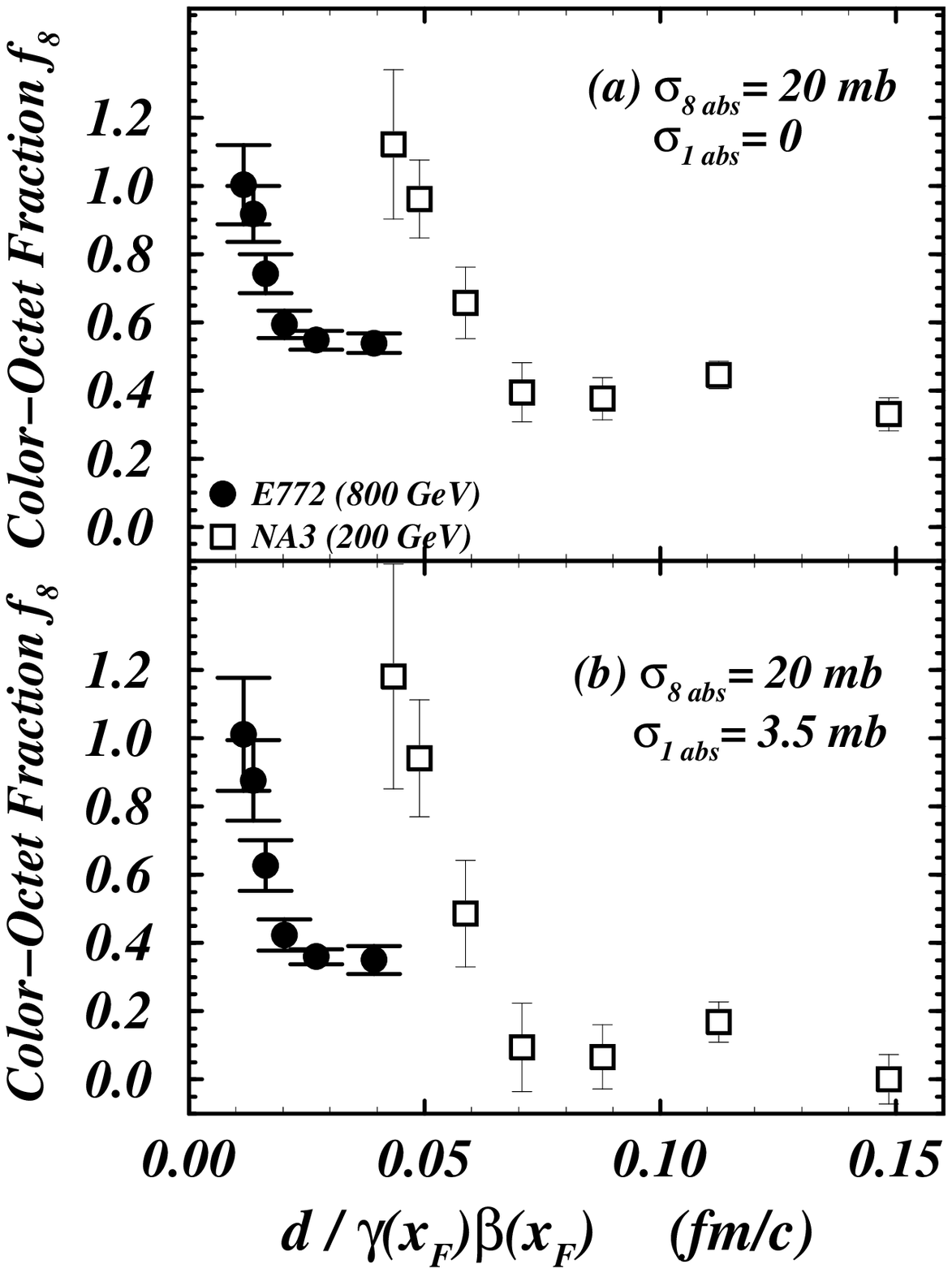}
\vskip 10.5cm
\begin{minipage}[t]{8cm}
\noindent {\bf Fig.4}.  \rm
{The color-octet fraction $f_8$ as a function of the passage
time $t_p = d/\gamma(x_{{}_F})\beta(x_{{}_F})$.  (a) is for
$\sigma_{\rm 1\, abs}=0$ mb , and (b) is for $\sigma_{\rm 1\,
abs}=3.5$ mb. }
\end{minipage}
\vskip 4truemm
\noindent 

The time scale involved in Fig. 4 is only a small fraction of the time
$1 / {\Lambda_{QCD}} \approx 0.5$ fm/c for nonperturbative color
neutralization by soft-gluon emission or absorption. The decrease of
C8 fraction with increasing passage time $t_p$ is particularly
noticeable in the model of Fig. 4b. It could be the consequence of a
relatively fast, or ``premature'', color neutralization mechanism
involving hard gluons.

\section{ Discussion and concluding remarks } 

A scenario rather close to premature color neutralization has been
proposed by Kharzeev and Satz\cite{Kha96}. They have suggested that by
the time the $(c\bar c)_8$ pair leaves the nucleon where it was
produced, its color has already been neutralized by the pickup of an
additional gluon to form a $J/\psi$ precursor that is a $(c\bar
c)_8-g$ hybrid \cite{Kha96}. Nuclear suppression comes from hybrid
absorption in hybrid-nucleon collisions. They estimate that the
required effective absorption cross section of about 6 mb for the
integrated yield is consistent with a hybrid size of size $r_8 \approx
1/\sqrt{2M\Lambda_{QCD}} \approx 0.2-0.25$ fm. It is not immediately
clear that this estimate of $r_8$ is theoretically reliable since it
is also an estimate for the size of the quark wave function in the
quarkonium. One would naively expect that with increasing quark mass
$M$, the hybrid size would decrease more slowly than the quarkonium
size, and perhaps not at all, if the effective gluon mass does not
change much with the quark mass $M$.

This hybrid picture has been used in \cite{Won96b} to interpret the
effective absorption cross section as a function of $x_{{}_F}$
obtained by us in a preliminary version of our analysis based on the
standard one-component analog of Eq. (\ref {eq:yratio}). In the TGMP
of hybrid-nucleon scattering, the total cross section is approximately
(9/4)$\sigma_{1 {\rm ~abs}}$ if the $(c\bar c)_8$ constituent is
treated as a point particle, and if the average $(c\bar c)_8$-$g$
separation is the same as that between the quark and the antiquark in
the $(c\bar c)_1$ quarkonium. However, the hybrid-nucleon cross
section can be made to vary by changing the hybrid size. The
$x_{{}_F}$ dependence of the deduced effective absorption cross
sections can then be translated into an $x_{{}_F}$ dependence of the
hybrid size, with the rms separation between the $(c\bar c)_8$ and $g$
ranging from about 0.14$\pm$0.02 fm for $x_{{}_F}=0.07$ to
0.5$\pm0.15$ fm at $x_{{}_F}=0.6$ \cite{Won96b}. The picture seems to
be that the gluon separation from the $c\bar c$ pair in the hybrid is
larger the higher the precursor energy.

The description given in Fig.\ 3 of the $x_{{}_F}$ dependence of
$J/\psi$ absorption in terms of a change in the C8 fraction also
differs from the explanation given by \cite{Gav92} based on the energy
loss of initial-state partons, the modification of initial target
parton momentum in nuclei, and the energy loss of final-state $c\bar
c$ systems.  The experimental open-charm production cross section in
$pA$ collisions has been found to behave as $A^{1.00\pm 0.05\pm 0.02}$
\cite{Alv93} for $x_{{}_F}$ from 0.05 to 0.4. Such a behavior implies
that the initial-state effects of energy loss and the modification of
initial target parton momentum distribution in nuclei are small.
Thus, the initial-state effects on $J/\psi$ production should also be
small.  Furthermore, the final-state precursor-$N$ collisions involved
are high-energy processes that are more likely to lead to eventual
breakup than to energy loss by quasi-elastic scattering.
 
Our absorption models have interesting implications in another aspect
of the nuclear absorption problem. The experimental nuclear
suppression of produced $\psi$' mesons appears to be quite similar to
that for $J/\psi$ mesons \cite{Lou95,Won96b}. For C1-dominated
absorption models, this could be understood only as a coupled-channel
effect, with different mesons propagating in nuclei in the same
coherent eigenmode and therefore the same ``eigen'' cross section
\cite{Won96b}. The need for coherent propagation is greatly reduced in
C8-dominated models, since the cross sections between C8 precursors
and nucleons are now size-insensitive, and about the same for C8
precursors of different $c$-$\bar c$ separations. Also cross-channel
matrix elements are likely to be quite small for Pomeron exchange
\cite{Won96b}. However, these C8 precursor channels could still be
coupled together via the exchanges of single hard gluons. Hence the C8
precursors for different quarkonium states might still propagate
together coherently.

In conclusion, we find that our absorption model with two color
components seems to be a useful tool for extracting the color-octet
fraction in quarkonium production under a variety of physically
interesting circumstances. The available experimental data on $J/\psi$
absorption in nuclei are not inconsistent with the theoretical picture
that color-octet precursors are abundantly produced and strongly
absorbed in nuclear collisions at fixed target energies, while
color-singlet precursors that are also produced might be transparent
because they are produced in pointlike states. However, better fits to
these data are obtained by using an older picture that color-singlet
precursors are significantly absorbed by nuclei and might be
responsible for most of the observed nuclear absorption by being
dominant in the absorption step of the reaction at least in certain
energy and kinematical regions.

Much more effort will be needed to clarify the situation.  It might be
necessary to have a better understanding and treatment also of
neglected effects in quarkonium production \cite{Tan96,Ben96}.  These
effects include higher-twist mechanisms of production \cite{Tan96},
higher Fock-space components in projectile or target
\cite{Bra96,Kha96}, and nonperturbative final-state interactions in
the production processes (the $K$-factor)
\cite{Fad88,Gus88,Cha95a,Sle96}, such as that between $c$ and $\bar c$
in color-octet production and between the $c\bar c$ pair and the
accompanying gluon in color-singlet production near threshold. In any
model, one needs to understand the nature and physical origin of the
$x_{{}_F}$ dependence of the extracted color-octet fraction. The
effect of the coherent mixing of precursor states at subsequent
collisions with nucleons should also be studied.

Above all, new experimental $J/\psi$ production data for different 
colliding nuclei at different energies will be very helpful in 
discriminating between different physical models, especially when 
extended to negative values of $x_{{}_F}$.

\acknowledgements

We would like to thank Drs.\ Yu-Qi Chen, C. S. Lam, and J. C. Peng for
helpful discussions.  This research was supported by the the Division
of Nuclear Physics, U.S.D.O.E.  under Contract DE-AC05-96OR22464
managed by Lockheed Martin Energy Research Corp.


\begin{references}
\bibitem{Foot1} 
Gupta and Sridhar \cite{Gup96,Gup96a} have not included the C1 
contribution from $gg\rightarrow J/\psi g$ in their analysis.  According 
to \cite{Tan96a} and \cite{Ben96}, this C1 contribution is about half of 
the C8 contribution from 
$gg\rightarrow {}^3P_J^{(8)} \rightarrow J/\psi g$, and should therefore 
be included. Because of this neglect, the estimate of \cite{Gup96,Gup96a} 
that $f_8$ is about 0.96 may have to be revised.
\end{references}
\end{document}